\documentstyle[aps,prl,preprint,epsfig]{revtex}
\tightenlines
\title{ Color Dynamics On Phase Space }
\author{A. Bonasera\footnote{Bonasera@lns.infn.it}}
\address{
Laboratorio Nazionale del Sud, Istituto Nazionale di Fisica Nucleare, 
Via S. Sofia 44, 95123 Catania, Italy $\&$\\
Cyclotron Institute, Texas $A\&M$ University, College Station, 
TX 77843-3366,USA.
 }
\begin{document}
\maketitle
\begin{abstract}
We describe the properties of quark matter at zero temperature 
and finite baryon densities within microscopic Vlasov/molecular
 dynamics approaches.
We use an inter-quark Richardson's potential
consistent with the indications of Lattice QCD calculations. The color
degrees of freedom are explicitly taken into account.  
 We explicitly demonstrate that the Vlasov approach alone is
insufficient in the hadronization region.  In order to overcome
this problem we prepare the initial condition for many events using molecular
 dynamics with frictional cooling and a Thomas-Fermi approximation to the
Fermi motion. 
  These events are averaged and propagated in time using the Vlasov
 approach.  
 We find  some evidence for a second order phase transition 
from nuclear to quark matter 
at high baryon densities.  An order parameter suitable to describe the
phase transition is discussed. At low densities the
quark condensate into approximately color white clusters (nucleon).
\end{abstract}


{\ \vskip 2\baselineskip

\newpage


One of the open problems in theoretical nuclear and particle physics is
how to obtain the well known nuclear properties starting from the
quark degrees of freedom \cite{pov}. This also includes the possibility of
understanding the basic free nucleon-nucleon interaction from quark and
gluons dynamics.  Some kind of solution to this problem is becoming more
and more needed with the new experiments done or planned using  
 ultra-relativistic heavy ions at CERN and soon at RHIC.  
The search for a quark-gluon plasma (QGP) in such collisions is in fact 
one of the new and most exciting
directions in physics at the border between nuclear and particle physics
\cite{wong}. Quantum ChromoDynamics
(QCD) because of its difficulties (numerical and conceptual),
has been applied so far to some
limited cases such as quark matter at zero baryon ($\rho_ B$)
density and high temperature
(T) \cite{pov,wong}. Furthermore in relativistic heavy ion collisions (RHIC)
dynamics plays surely an important role and accordingly the theory should 
 be dynamical.

 Recently \cite{bon99}, 
we have proposed a dynamical approach based on the
Vlasov equation \cite{repo,land1} to reproduce hadron masses and the properties
of nuclear matter at finite $\rho_B$. 
Some works in the same spirit are  discussed in \cite{mosel,horst}.
Our approach
needs as inputs the interaction potential among quarks, which was borrowed
from phenomenology i.e. the Richardson's potential\cite{rich}, and the quark 
masses which were fitted to reproduce 
known meson masses such as the $\pi$, the $\phi$, the 
$\eta_c$ etc.  When the particles are embedded in a dense medium such as
in nuclear matter (NM) the potential becomes screened in a similar fashion
as ions and electrons in condensed matter do, i.e. 
Debye screening (DS)\cite{wong,land1}. 

It is the purpose of this paper to refine that approach in one important
aspect which is the treatment of the color degrees of freedom.  In the
previous works \cite{bon99} color degrees of freedom were implicitly
taken into account through the use of a Debye radius that effectively
screens the $qq$ interaction potential.  In the present paper we 
 give to the quarks explicitly a color (using the Gell-Mann matrices)
and follow their dynamics in phase space solving the Vlasov equation (VE).
 Thus screening will be dynamically obtained.  
In general the self screening
 obtained in Vlasov dynamics is inadequate, which is the reason why it was not
adopted in the earlier attempts \cite{bon99}.  In fact we will
 show explicitly that the Vlasov approach alone gives a good 
description of the system at large densities only,i.e. in the
 QGP region.  In order to overcome this
problem we adopted the following strategy. 
We first prepare the initial conditions using molecular dynamics (MD)
with frictional cooling for many events.  The events are averaged
and care is taken of antisymmetrization. These are the initial conditions
for the Vlasov evolution.  Since the VE fulfills the Liouville theorem,
the initial phase space density remains constant in time. Thus the initial
antisymmetrization and eventual clustering obtained in the cooling process
are maintained.

We outline our approach on purely classical grounds, however
the same results can be obtained within the Wigner transform formalism
\cite{repo,land1} of the quantum BBGKY-hierarchy in 
the limit $\hbar \rightarrow 0$.

The exact (classical) one-body distribution function $f_1(r,p,t)$ 
 satisfies
the equation (BBGKY hierarchy)\cite{land1}:
   \begin{equation}
\partial_{t}f_1+\frac{\overrightarrow{\text{p}}}{E}\cdot \nabla _{r}f_1
=\int{d(2)\nabla_r V({\bf r},{\bf r_2})\nabla_p f_2({\bf r,r_2,p,p_2},t)}
\label{lv1}
\end{equation}
$E=\sqrt{p^2+m_i^2}$ is the energy and 
$m_i=10 MeV$ is the (u,d) quark mass. 
Here we assume the potential to be dependent on the relative coordinates only.
A generalization to include a momentum dependent part is straightforward.
 $f_2$ is the two-body distribution function, which in the classical limit
reads:
\begin{equation}
f_2(r,r_2,p,p_2,t)=\Sigma_{\alpha \ne \beta}^{Q}\delta ( \bf r- \bf r_{\alpha})
\delta (\bf p- \bf p_{\alpha}) \times \delta ( \bf r_2- \bf r_{\beta})
\delta ( \bf p_2- \bf p_{\beta})
\label{lv2}
\end{equation}
where $Q=q+\bar q$ is the total number of quarks and anti quarks ($\bar q=0$
in this work).  
Inserting this equation into Eq.(1) gives:
\begin{equation}
\partial_{t}f_1+\frac{\overrightarrow{\text{p}}}{E}\cdot \nabla _{r}f_1
-\nabla_{\text{r}}U\cdot \nabla _{\text{p}}f_1=0
\label{lv3}
\end{equation}

Where $U=\Sigma_j V(\bf r,\bf r_j)$ is the exact potential.  Let us now
define $f_1$ and $U$ as sums of an ensemble averaged quantity plus the
deviation from this average:
\begin{equation}
f_1=\bar f_1+\delta f_1;  
U=\bar U+\delta U
\label{lv4}
\end{equation}
Substituting these equations in Eq.(3) and ensemble averaging gives:
   \begin{equation}
\partial_{t}\bar f_1+\frac{\overrightarrow{\text{p}}}{E}\cdot 
\nabla _{r}\bar f_1
-\nabla
_{\text{r}}\bar U\cdot \nabla _{\text{p}}\bar f_1= 
<\nabla_r\delta U\nabla_p\delta f_1>
\label{lv}
\end{equation}

where one recognizes in the lhs the Vlasov term and in the rhs the
Balescu-Lennard collision term \cite{land1,belk}.
The mean-field is given by:
    \begin{equation}
\bar U(\bf r)=\frac{1}{N_{ev}}\Sigma_{ev}\Sigma_j V(\bf r,\bf r_j)
\end{equation}
For the purpose of this work we neglect the collision term in Eq.(5)
 and note that such term will be essential when dealing with RHIC.
In agreement to LQCD calculations\cite{pov,rich} the interacting potential
$V(r)$ for quarks is ($\hbar=1$):
 
 \begin{eqnarray}
 V(r_{i,j})=3\Sigma_{a=1}^{8}\frac{\lambda_i^a}{2}\frac{\lambda_j^a}{2}[
\frac{8\pi}{33-2n_f}\Lambda(\Lambda r_{ij}-\frac{f(\Lambda r_{ij})}
{\Lambda r_{ij}})
+\frac{8\pi}{9}
\bar{\alpha}\frac{<{\bf \sigma}_q{\bf \sigma}_{\bar q}>}
{m_q m_{\bar q}}\delta({\bf r_{ij}}) ]
 \end{eqnarray}
 
 and\cite{rich}
 \begin{eqnarray}
f(t)=1-4 \int{\frac{dq}{q}\frac{e^{-qt}}{[ln(q^2-1)]^2+\pi^2}} 
 \end{eqnarray}

We fix the number of flavors 
$n_f=2$ and
the parameter $\Lambda=0.25 GeV$ 
.
In Eq.(7) we have added to the Richardson's potential 
the chromomagnetic term (ct), very important
 to reproduce  the masses of
the hadrons in vacuum.  
Since in this work we will be dealing with finite nuclei, the ct can
 be neglected, we only notice that with the parameters choice discussed here,
the hadron masses can be reproduced by suitably tuning the ct term
\cite{bon99}. 

  The $\lambda^a$ are the Gell-Mann matrices.
  From lattice calculations we expect that there is no color transport for
distances of the order of $0.2-0.3 fm$, which are distances much shorter
than the ones we will be dealing with in this paper. Thus we will use
 the $\lambda_{3,8}$ only commuting diagonal Gell-Mann matrices (Abelian
 approximation)\cite{horst}.
 
Numerically the VE equation(5) is solved by writing the one body distribution
function as:

\begin{equation}
\label{efer}
\bar f_1(r,p,t) = \frac{1}{n_{tp}} \sum_i^N \delta(r-r_i(t)) \delta(p-p_i(t)) 
\end{equation}

 $N=Qn_{tp}$ is the number of such terms.
  Actually, 
N is much larger than the total quark
number Q, so that we can say that each quark 
is represented by ${n_{tp}}$ terms called test particles(tp).
  Notice
how well this fits in the previous discussion if we put $n_{tp}=N_{ev}$.
 Inserting Eq.(9) in the Vlasov equation
 gives
the Hamilton equations of motion (eom) for the tp~\cite{repo}.
%
%
The total number of tp (or corresponding $N/Q$ events)used in this work
ranges from 5000 to 50000, and  Q=150-300. 

Initially, 
 we distribute randomly  the tp in a sphere of radius $R=r_{0B}A^{1/3}$ 
(the radius of the nucleus) in coordinate
space and $p_f$ in momentum space.  $r_{0B}=(\frac{3}{4\pi\rho_B})^{1/3}$,
$A=Q/3$ and $\rho_B$ is the baryon density.
$p_f$ is the Fermi momentum
estimated in a simple Fermi gas model by imposing that a cell in
phase space of size $h=2 \pi$ can accommodate at most two identical quarks
 of different spins, flavors and colors. A
simple estimate gives the following relation between the quark density
 ,
$n_{q}$, and the Fermi momentum:
\begin{eqnarray}
n_{q}=\frac{g_q}{6\pi^2}p_f^3 
 \end{eqnarray}
  The degeneracy number 
$g_q=n_f\times n_c\times n_s$, where $n_c$ is the number of colors and
$n_s$ is the number of spins\cite{wong}.  For quarks and
 anti quarks 3 different colors are used red, 
yellow and blue (r,y,b) \cite{pov}.

In figure(1), we plot the total energy per nucleon (top) and energy
 density (in units of the Fermi gas energy density\cite{wong}) vs.
baryon density. The full triangles give the results
obtained by randomly distributing the tp as described above.  We notice
that  a minimum at about $\rho_c=2.08 fm^{-3}$ is found with $E_t/A=2GeV/A$.  
Such a minimum is at much higher density and energy than expected for
the ground state (gs) of a nucleus 
($\rho_0=0.16 fm^{-3}$ and $E_t/A=0.938-0.016 GeV/A$).
An important property of the system that we have  described
above is the following.  If we rotate the quarks in color space,
regardless of their position in r-space, the
total energy will remain the same. 
This is indeed a "pure" Vlasov solution and demonstrates explicitly the
in-capability of the Vlasov approach to give clustering of quarks into
 nucleons.  However this result is already instructive since it gives us an
 hint on where the quark and nuclear matter are located, i.e. above and below
$\rho_c$ respectively. This result is qualitatively in agreement
with the Hartree-Fock(HF) calculations of refs.\cite{horo,roe}
(compare to fig.1 in \cite{horo}).
For a discussion on why the $E_t/A$ increases at low densities in the HF/VE
 approaches we also refer to \cite{horo}. 

  Of course distributing randomly
the quarks in a sphere in r and p-space is not the most economical way
 to find the real gs of the system.  In MD one 
searches for a minimum energy by introducing a friction term.  
  The friction acts in such a way to lead the
 particles to a configuration for which the potential energy is a minimum.
 We cannot use this technique for our system since we are dealing with
 fermions and the friction term will destroy the initial 
antisymmetrization.  In order to overcome this difficulty we adopt the
following strategy.  First we prepare $N_{ev}$ events as above, and
we evolve them numerically solving the  
eom but with  friction included. 
Because of the large number of particles interacting with attractive 
and repulsive forces, the quarks will slowly evolve to new
 positions where the potential is lower while keeping the initial
root mean square radius approximately constant.  When the averaged potential 
(over events) reaches a given value $V_{min}$, 
we look for the two closest particles $\it (j,k)$ to
a quark $\it i$ in the same event.  
For these three quarks we know what the local
density and the number of colors are.  For instance if in a certain region
we find two red and one blue quark,  we use $n_c=2$ in Eq.(10) and
calculate 
the local density from the knowledge of the distances of the 3 quarks.  
In this way the Fermi momentum is defined locally
similar to the procedure used in nuclear or atomic physics (Thomas-Fermi
approximation)\cite{schuck}. We repeat this for all quarks $\it i$
 in all events and calculate
 the total energy for this state
.  We let the system evolve again with friction included 
to a lower potential energy $V_{min}=V_{min}-\delta V$, where $\delta V$ is a 
constant. We calculate the local density and local color numbers again
and apply the Thomas-Fermi approximation to obtain the new total energy.  
We stop the procedure when
the total energy is a minimum. The initial conditions so obtained are
 then propagated in time using the VE in order to maintain the initial
 antisymmetrization. We will show below, see fig.3, that the initial
clustering obtained in MD is preserved as well during the Vlasov evolution.  

The open triangles in fig.(1) are the
result of the minimization.  Notice especially at low densities the
large decrease of the total energy of the system as compared to the
Vlasov result.  Now the calculated 
total energy
at the nuclear gs is very close to the experimental value indicated by the
full circle. However, we find slightly lower energies at lower densities, 
 i.e. the gs of the nucleus is shifted in our calculations at about
1/10 of the experimental one.  This should not be surprising in view of 
the simple potential that we have used. Also we have not tried to
look for a best set of fitting parameters in these exploratory studies.  
At low $\rho_B$,
 the global invariance for rotations in color space
is lost, i.e. if we exchange the colors of two ${\it distant}$ quarks, 
the total
energy of the system will change.  
  At high densities (larger
than $2 fm^{-3}$) 
the Vlasov and MD solutions are the same.
  This can be also seen in the bottom part of the figure where energy
densities are given. We would like to stress again the qualitative agreement 
to ref.\cite{horo} where a stochastic method had been used to calculate the
g.s. energy of the system.  This is quite evident if one compares our fig.1
 to fig.1 of \cite{horo}.

The energy density displayed in fig.1 (bottom) is a smooth
function apart some fluctuations around $2 fm^{-3}$ density.  From
this result we can exclude a first order phase transition but a second
order phase transition might be possible.
   
In order to check if a second order 
phase transition occurs we define an order parameter.
As we discussed above the color degrees of freedom play an important role for
our system.  When the quarks of different colors are in suitable positions
in r-space the potential energy largely decreases.  It is also
very important that the system is locally color white because in this way
$n_c=3$ and the kinetic part will also decrease.  Thus for extremely
small densities the quarks should condensate in clusters of 3  and
zero net color.  When the density changes this picture gradually modifies
 and at very high densities it does not matter where the
quarks are located and which color they have.  Looking at Eq.(7) we see
that this fact is a consequence of the 
$\frac{3}{4}\sum_{a=3,8} \lambda_i^a \lambda_j^a$
term (equal to $-1$ for identical color quarks and $1/2$ otherwise)
.  Thus we  
 define an order parameter $M_c$ as:
\begin{equation}
M_c
=\frac{1}{N}\sum_{i=1}^{N}\sum_{a=3,8}(\lambda_j^a\lambda_k^a+
\lambda_i^a\lambda_j^a+\lambda_i^a\lambda_k^a)=
M_{c_r}+\frac{1}{N}\sum_{i=1}^{N}\sum_{a=3,8}(\lambda_j^a\lambda_k^a+
+\lambda_i^a\lambda_k^a)
\end{equation}

Where $j,k$ are the two quarks closest in r-space to quark $\it i$ as
before.  In Eq.(11) we have also defined a "reduced" order parameter
$M_{c_r}$ which tells us the color of the closest particle $j$ to
quark $i$.  From the
properties of the (3,8) Gell-Mann matrices it is easy to derive the following
  results for $M_c$. 
 If the 3 closest quarks have  
the same color, $M_c=-3$.
We stress that this case is practically impossible to be obtained because
the corresponding potential energy would be very large and repulsive.
If the 3 closest quark states have two colors $M_c=0$, this case is also
recovered if the colors are randomly distributed such as in the "pure"
Vlasov solution.
The case of 3 different color quarks gives 
$M_c=3/2$.  The last is the ideal case of well isolated white nucleons.  If
this last case is recovered in the calculations at small densities
then the system is ${\it locally}$ 
invariant for rotation in color space.  i.e.
if we rotate the color states inside the nucleon, the total energy of the
system will remain constant.  Using similar arguments it is simple to show
that if the closest particle to quark $i$ has always 
a different color 
$M_{c_r}=1/2$, if the two closest quarks have the same color
 $M_{c_r}=-1$.  If the closest quark color 
is randomly chosen $M_{c_r}=0$.

In figure (2) top , we plot the order parameter 
(opportunely normalized) vs.
density (divided by a critical density-see below) 
for the MD case. 
The displayed $M_c$ are 
 always positive i.e. it never happens that 3 equal
quark color states are on average in the same region in r-space.  
In the top part of the figure we have distinguished two cases.  The first
one indicated by the full triangles corresponds to calculations where the
average potential energy is larger than zero i.e. the linear term in
Eq.(7) is dominant, small $\rho_B$.  The full
 squares correspond to the case where the Coulomb term 
is dominant, large $\rho_B$ and negative mean field.  
The first case can be rather well fitted by the relation:
\begin{equation}
M_c \propto |1-\frac{\rho_B}{\rho_c}|^{\beta}
\end{equation}
Where $\rho_c=2.08 fm^{-3}$ is the critical density and the two curves
correspond to the critical exponent values $\beta=1/3$(full line) and $1/2$
(dashed line).  The latter is the expected value of the critical
 exponent $\beta$ in a mean field approach \cite{huang}
.   We notice that in ref.\cite{roe} a phase transition of
first order was found.  However the potential and the kinetic term
(non-relativistic) used there are quite different from ours.  In  LQCD 
calculations for Fermions at zero temperature the order of the phase
transition depends on the quark masses. For small quark masses such
 as ours, the transition found 
is second order \cite{wong,bro}.  

In order to better understand the behavior of $M_c$ we
have repeated the calculations by turning off the Coulomb term (open triangles)
or the linear part (open squares).
At low densities
the linear term is larger while the Coulomb term is dominant at high 
densities. The two terms are equally important around the critical density. 
 However, there is an important difference between the two cases.  In fact
at low densities the kinetic part is rather small compared to the 
potential one, while it is rather large
at high densities, Eq.(10).  
Furthermore, at high densities the strong coupling constant
entering the Coulomb potential, Eq.(8), vanishes logarithmically\cite{rich}.
  Thus it is
clear that the order parameter increases again at high densities
because of the friction used to  lower the potential energy.  But as soon as
we "turn on" the kinetic term we expect the bonds to be quickly broken.  This
is shown in figure (3) where the time evolution of $M_c$ is given.  At high
density we have a large value of $M_c$ (full circles) at time t=0fm/c.  But
such a value quickly decreases to a minimum value of about 0.3.  The $M_c$
obtained at low densities is rather large and constant (open circles),
 while the one obtained at the critical density slightly increases (open 
squares).
This result also shows that the VE can keep the initial clustering
obtained from the MD initial conditions for a time sufficient to perform
calculations for heavy ion collisions at relativistic energies. 
It also implies that the $M_c$ values obtained at high $\rho_B$ are an
artifact of the calculation and we expect $M_c$ to saturate at about
0.3-0.4 at high densities.  This is exactly the behavior expected for a
order parameter, i.e. a power law dependence for densities below the critical
one and constant otherwise.  It is also instructive to notice that the
two closest quarks have always a different color for low densities as
it is suggested by $M_{c_r}$ in the bottom part of figure(2).  
The $M_{c_r}$ starts to be smaller than 1/2 for densities
larger than $\rho_0$ and approaches zero very slowly with increasing
density. 

From figures (2) and (3) some important consequences can be derived.

i) The order parameter is never equal to 3/2 i.e. 
isolated white nucleons.
  As it is shown in figure 2, $M_c$ is still increasing
with decreasing densities.  Thus in the limit of very small densities
we should get color white objects.  However such a limit is hard to reach
 because of numerical fluctuations due to the confining 
 potential.
The maximum calculated $M_c$ value is  about
0.9,a value in between 2 and 3 color states. 
Of course it is not always the same cluster (nucleon) to have $n_c=2$
or 3, but
rather the number of colors in a cluster changes dynamically between 2 and 3.
 In order to understand this behavior imagine to have two clusters
with one color exchanged, $(r_1,y_1,y_2)$ and $(r_2,b_1,b_2)$, 
located at very large distances.  The contribution to the potential of cluster
1 due to cluster 2 is zero for the red quark (-1+1/2+1/2) and 3/2 for each
of the yellow quarks.  Thus the two cluster will be $\it attracted$ towards
each other at low densities.  On the other hand there is a repulsion
between the two equal color quarks in the two clusters.  These
 quarks will be pushed away  from their original cluster and eventually  
the white color will be established. In other words the "color migration" 
   binds the clusters.

ii) The system is not locally ($M_c=3/2$) 
nor globally ($M_c=0$) color invariant.
 This is true at high densities as well, but there the potential
energy is negligible as compared to the kinetic one and color invariance
is (approximately) restored. 

In conclusion in this work we have discussed  microscopic Vlasov/MD 
approaches
to finite nuclei starting from quark degrees of freedom with colors.
In order to obtain the correct initial conditions we have introduced a method
based on MD with frictional cooling plus a Thomas-Fermi approximation for
the Fermi motion.  
We have shown that the method is able to describe at least qualitatively
 the well known features of nuclei near the ground state.  At high
densities a second order phase transition from nuclear to quark matter
is predicted.  Such a transition is due to the restoration of global
color invariance at high densities and we have defined an order parameter
accordingly. The approach can be refined in order to obtain a better 
description of the ground state of the nucleus.  This can be used to 
simulate heavy ion collisions at ultra-relativistic energies after 
the introduction of a suitable collision term.  Our approach can be
very useful for the understanding of the quark gluon plasma formation and
its signatures.

 {\ \vskip 0.7 cm \centerline{\bf ACKNOWLEDGMENTS} }
We thank prof. J.B. Natowitz and the colleagues at  the Cyclotron 
Institute-Texas $A\&M$ University
 for warm hospitality and financial support.



%
%
%
%
\begin{figure}[tbp]
\begin{center}
\mbox{{ \epsfysize=14 truecm \epsfbox{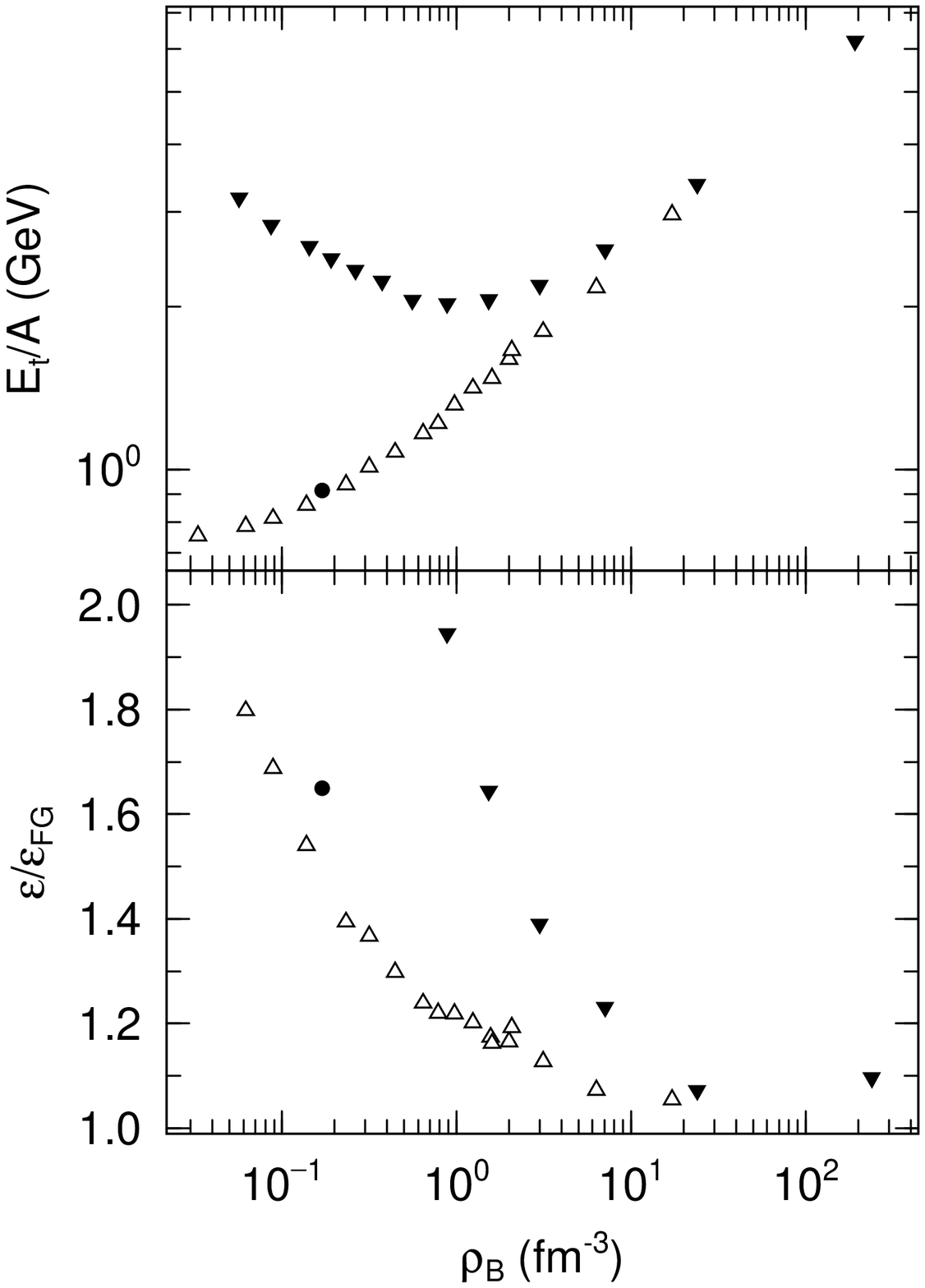}}}
\vskip 1.5cm
\caption{Energy per nucleon (top) and energy density
of the quarks 
(bottom) vs. baryon density. The full triangles refer to Vlasov 
and the open ones to the molecular dynamics initializations (see text). 
The full circles give the
values of the nuclear matter ground state. 
\label{fig:fig1}}
\end{center}
\end{figure}
\begin{figure}[tbp]
\begin{center}
\mbox{{ \epsfysize=12 truecm \epsfbox{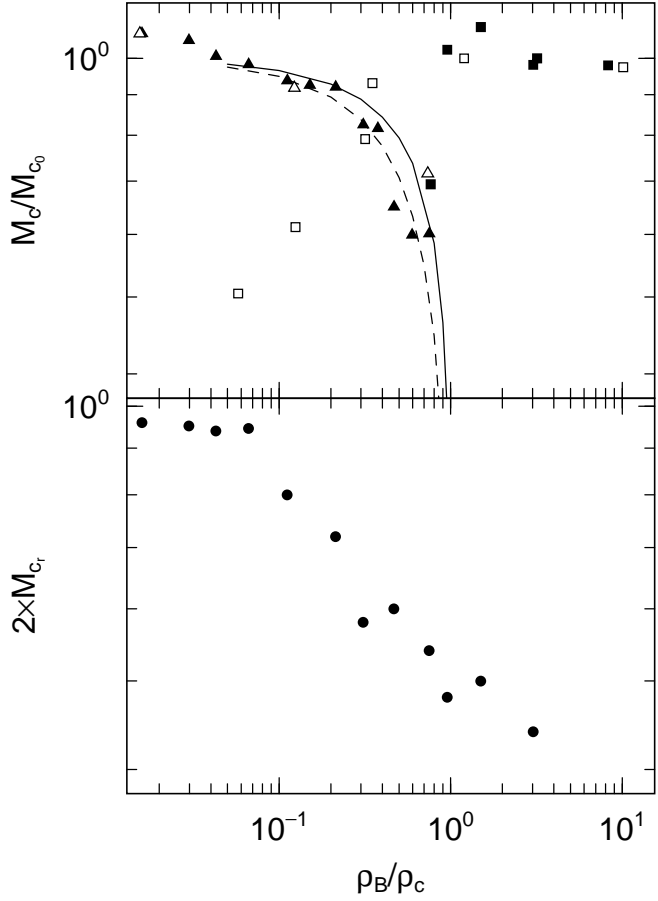}}}
\vskip 1.5cm
\caption{
Order parameter(top) and "reduced" order parameter (bottom)
 vs. $\frac{\rho_B}{\rho_c}$. Critical density
$\rho_c=2.08 fm^{-3}$.  The top part refers to the full calculation when
the average potential is larger than zero (full triangles) or less than
zero (full squares). The lines are given by  $|1-\frac{\rho_B}{\rho_c}
|^\beta$ with $\beta=1/3$ (full line) or $\beta=1/2$ (dashed line).
 Also calculations are reported where the linear part of the
potential is turned off (open squares), or the Coulomb part is off (open
triangles).
 \label{fig:fig2}}
\end{center}
\end{figure}
\begin{figure}[tbp]
\begin{center}
\mbox{{ \epsfysize=12 truecm \epsfbox{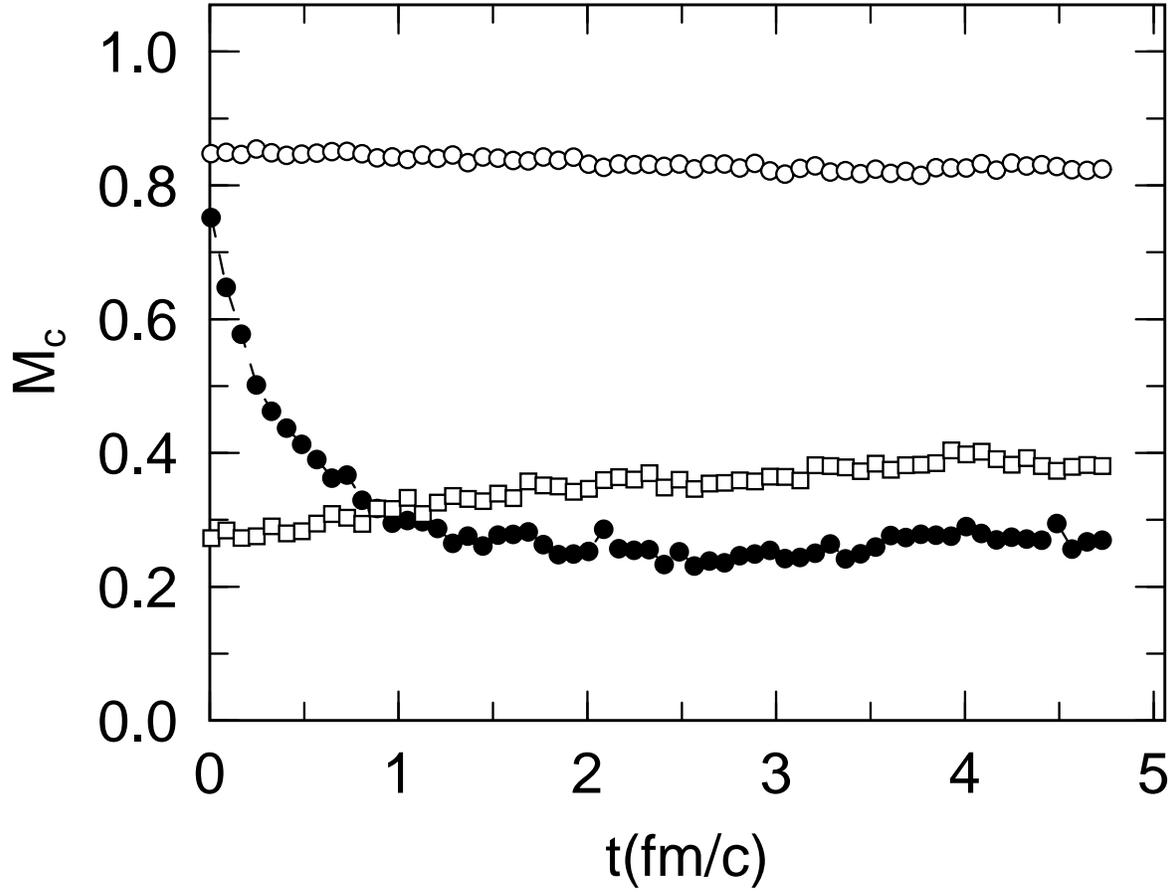}}}
\vskip 1.5cm
\caption{ Order parameter vs. time calculated in molecular dynamics
 plus Vlasov approach at three initial densities: 
$0.033 fm^{-3}$ (open circles), $29.8 fm^{-3}$ (full circles), and
 $2.08 fm^{-3}$ (open squares).
\label{fig:fig3}}
\end{center}
\end{figure}

\end{document}